# From Omicron to Its BA.5 Subvariant:

# A New Twist in Coronavirus Contagiousness Evolution

J. C. Phillips

Dept. of Physics and Astronomy

Rutgers University, Piscataway, N. J., 08854

Abstract

In 2022 new subvariants of Coronavirus appeared in South Africa, and spread rapidly to other parts of the world. The effects of selective evolution on increasing the already high contagiousness of CoV19 up to Omicron were previously discussed with high accuracy using concepts new to molecular biology (from mathematics and physics) that have been developed over the last few decades. Here we discuss the new subvariants, and find novel features successfully combining both improved flexibility and improved stability.

Introduction

Phase transition theory [1-4], implemented quantitatively by thermodynamic scaling [5,6], has explained the evolution of Coronavirus' extremely high contagiousness caused by a few key mutations from CoV2003 to CoV2019 identified among hundreds, as well as the later evolution to Omicron [7-9] caused by 30 mutations. Recently CoV subvariants, labelled BA.N, have emerged, and one of them (N = 5) has spread very rapidly in Portugal and the United States. It has spread less rapidly in South Africa, where it originated, and where N = 4 is still dominant [10]. This behavior is unusual, and will be discussed below.

Our earlier analysis relied on Ψ(aa,W) hydropathic profiles, where Ψ(aa) measures the fractal hydropathicity of each amino acid [5,6]. These have been averaged linearly over sliding windows of width W. This width was chosen initially to maximize evolutionary differences. Studies of many proteins then showed that evolution often resulted in level sets of extrema of Ψ(aa,W). Level sets have been used by James Sethian, an applied mathematician, to study interface propagation and assembly in many systems, including fluid mechanics, semiconductor manufacturing, industrial inkjets and jetting devices, shape recovery in medicine, and medical and biomedical imaging [11]. He also uses the ancient Voronoi topological construction to separate and organize interfaces [12]. These multiple and diverse applications suggested that level sets can also be useful in quantifying evolutionary trends in protein dynamics.

Results

The shape of a globular protein depends strongly on its ins and outs relative to its center. The limits of these ins and outs are associated with hydrophobic and hydrophilic extrema. The shape of the protein changes during phase transitions. A given protein can function mainly through either first- or second-order phase transitions, and second order is more common [3,4,7]. Protein unfolding from water to air is a first-order transition, well-described by the most popular hydropathic scale (KD) [13]. Second-order thermodynamic transitions involve much smaller structural (long-range or allosteric) transitions, such as may occur in viral cellular attachment and penetration [7].

CoV evolution has occurred very rapidly, and linearized sliding windows are only a first approximation for describing dynamical domain synchronization. The mutations of BA.2 and BA.5 are concentrated in the receptor domain region [13]. Here we will find the emergence of a modified less level pattern, with the domain synchronization found earlier for Omicron evolving more in BA.5 and less in BA.2. The sequence mutations are reported in [14]; the spike sequences are identical in BA.4 and BA.5. We continue to use $W = 39$ [8], with results shown in Figs. 1 and 2. Most of the edge shifts in the subvariants occur in a dynamically extended static receptor domain region. Thus we focus on the edges numbered in Figs. 1 and 2, with the same numbering as used for Omicron [9]. The differences in edges can be seen in these figures, but as they are small (~ 1% of the profile range) the numbers are shown in Table 1.

The largest differences are in two edges, 1 and 8, in the RCB, closely followed by nearby 7. They increase from KA.2 to KA.5. At first these trends seem to run counter to the effects of synchronization by edge leveling. However, the deep hydrophilic minimum 1 represents the outer edge of the spike, which will now reach its exterior contact site on ACE2 [15] faster. The increased hydrophobicity of edge 8 increases the interior stability during folding of the spike as it binds. Although edge 7 nominally lies outside the static RCB, it is nearly level with edge 8, and can synchronize with it during the folding process. Evolution of edge 7 increases its interior stability during folding of the spike hydrophobicity in tandem with edge 8.

Is there another nonlinear effect that contributes to increasing contagiousness of KA.5? Synchronization is important, and to a lesser extent domain flexibility can speed attachment. Broadly speaking, spike proteins are divided into two nearly equal parts, S-1 and S-2. The S-2 parts are involved in infection by fusion, while the RCB is located in the S-1 part. Stiffening the S2 part quickly produced highly effective vaccines. Initially this was accomplished by making two adjacent proline substitutions (V976P and L977P) in the center of the longest helix of S-2 (913-1032, Uniprot P0DTC2), which increased yield by a factor of 50 [16]. Later smaller improvements were accomplished mainly by substituting pairs of prolines at nearby sites [17]. A recent study observed a strong trimeric stabilizing effect with triplet proline substitutes T941P, A942P, and A944P [18]. Proline is the only amino acid which is connected twice to the peptide backbone, and proline pairs have often been used to stiffen protein chains. In [17] it was found that disulfide bonds were ineffective stabilizers because of "interference", here explained specifically by interference of interdomain water waves [9].

There are three PP pairs in CoV-2; two are in S-2. The third pair is in S-1 in the N terminal domain (NTD), and it is deleted in BA.(4,5), 25-27PPA. Note here that alanine is the smallest protein amino acid, making the original triad less crowded. When the PP pair was deleted, it seems natural that it took its neighbor A as well. Three of the four later best vaccine proline replacements also were A by P [17]. Just as adding PP pairs increases stability, deleting 25-27PPA makes the BA.(4,5) NTD more flexible.

Discussion

Spike proteins (1200 amino acids) are long and slender and immersed in water. When natural selection engineered the evolution of CoV-1 into CoV-2 with 300 mutations, there was a very large increase in contagiousness. Simultaneously the mutations stabilized trimeric fusion, making the virus much more dangerous. Our earlier analysis [7-9] has shown how the increase in leveling of extrema of Ψ(aa,W) correlates well with increasing contagiousness up to and including Omicron. Note that rapid attachment promotes localization of the virus in the upper respiratory tract [19]. From there it can be expelled to aerosols, and rapidly spread. At that point it might have seemed that there was little room for further evolution, but there remained still many possible mutations of Ψ(aa,W).

Here we have found that these involve not only improvements in the RCB hydropathic extremeties, but also a more flexible NTD, achieved by the PPA deletion. Altogether there are two BA.(4,5) features that are made more stable by evolution, the RCB hydrophobic edges (7,8) and two features that are made more flexible, hydrophilic edge 1, and the PPA deletion. Normally proteins are modified to make them either more flexible or more stable, but not both [20,21]. Evolution has made BA.(4,5) dynamics both more stable and more flexible topologically. This indicates the closest approach to self-organized criticality [1] known to date, accessible only by thermodynamic scaling [3].

As for the large national differences in spreading of BA.5 [10], to explain how it can occur first in South Africa, but spread most rapidly elsewhere (especially Portugal and the United States), these could be explained by two factors, such as climate and diet.

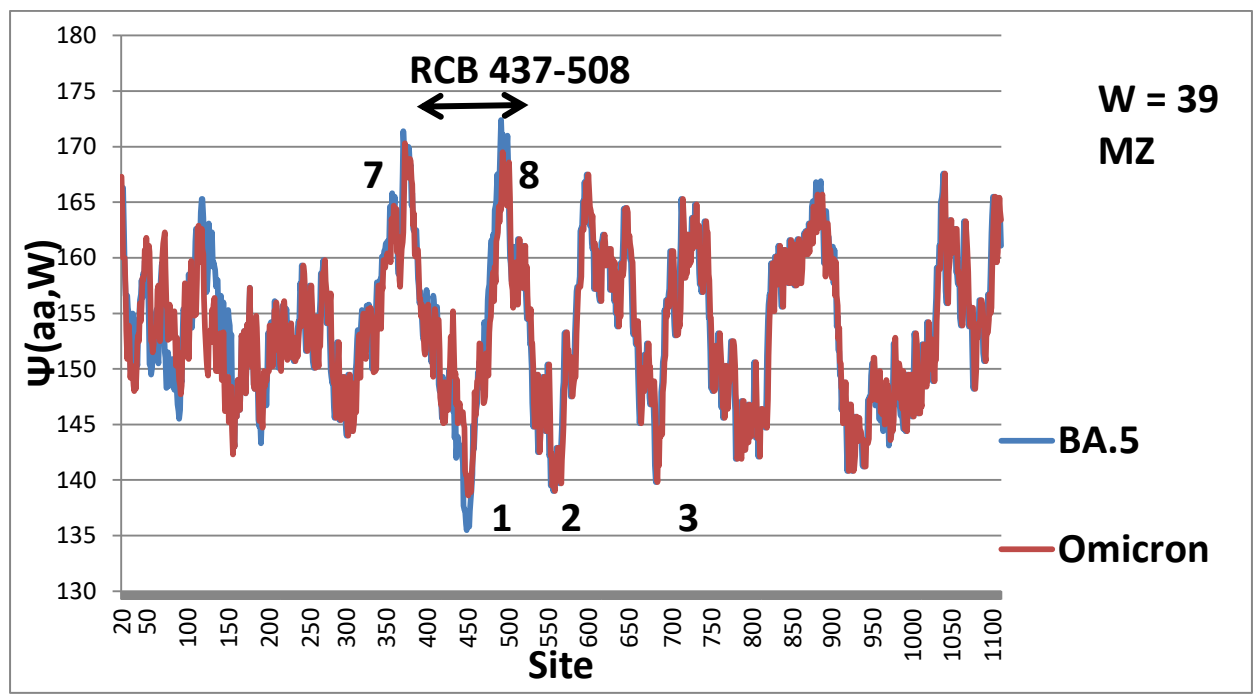

Fig. 1. The hydropathic wave profiles of Omicron and BA.5 are very similar, and the most important changes are in edges 7,8 and 1. See table 1; data from [14].

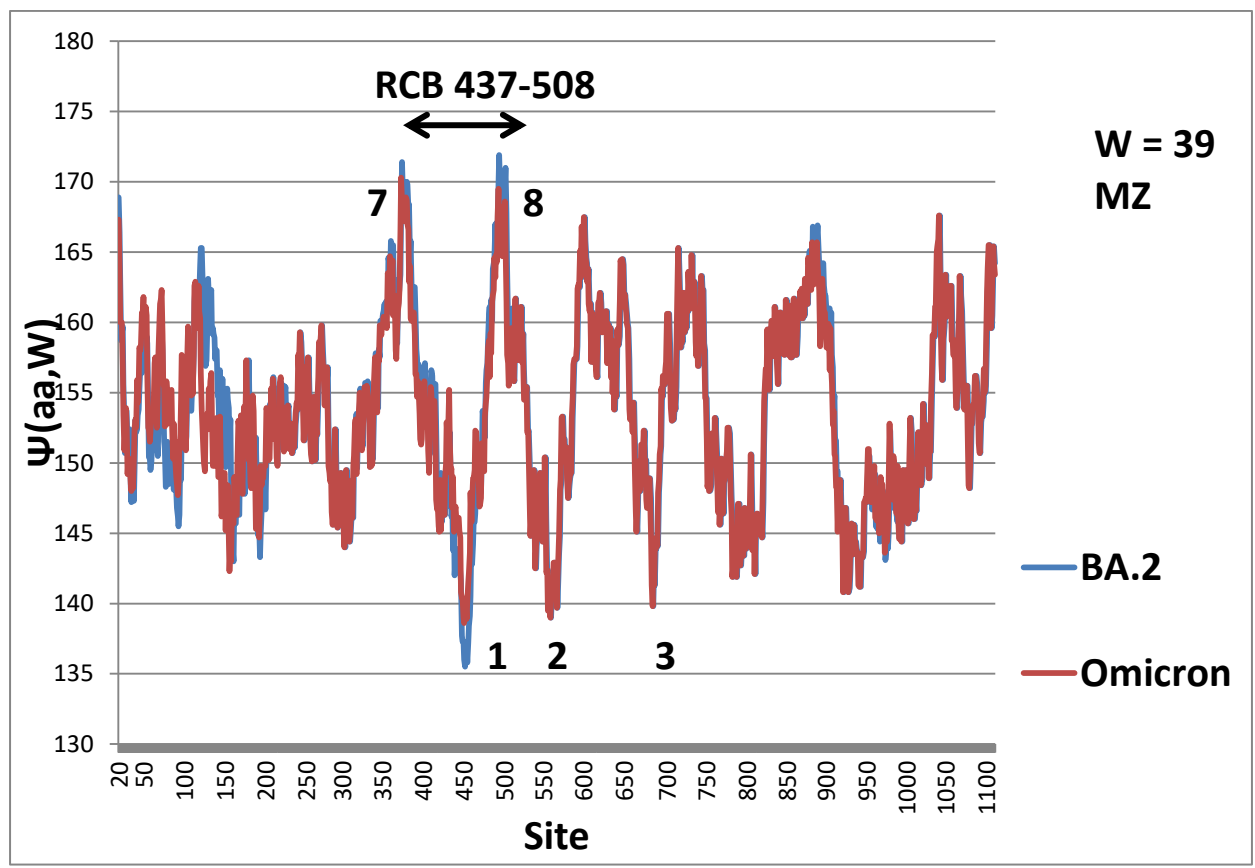

Fig. 2. As in Fig. 1, here for Omicron and less evolved BA.2

| Edge | Omicron | BA.5 | BA.2 |
|---|---|---|---|
| 7 | 170.3 | 171.4 | 171.4 |
| 8 | 169.4 | 172.4 | 171.9 |
| 1 | 138.6 | 135.5 | 135.5 |
| 2 | 139 | 139 | 139 |
| 3 | 139.8 | 139.8 | 139.8 |

Table 1. Edges (extrema) values from Figs. 1 and 2. The subvariant (7,8,1) edge shifts are small but systematic in increasing contagiousness, as discussed in the text.

Table 1. Extrema (edges)